# Next generation Co-Packaged Optics Technology to Train & Run Generative AI Models in Data Centers and other computing applications

John Knickerbocker[4], Jean Benoit Heroux[1], Griselda Bonilla[1], Hsiang Hsu[3], Neng Liu[3], Adrian Paz Ramos[1], Francois Arguin[1], Yan Tribodeau[1], Badr Terjani[1], Mark Schultz[4], Raghu Kiran Ganti[3], Linsong Chu[3], Chinami Marushima[2], Yoichi Taira[2], Sayuri Kohara[2], Akihiro Horibe[2], Hiroyuki Mori[2] and Hidetoshi Numata[2]

[1]IBM Bromont, 23 boul. de l'aeroport, Bromont, QC J2L 1A3, Canada
[2]IBM Japan, 7-7 Shin-Kawasaki, Saiwai-ku, Kawasaki 212-0032, Japan
[3]IBM Albany, 257 Fuller Road, Albany, NY 12203
[4]IBM T. J. Watson Research Center, 1101 Kitchawan Rd, Yorktown Heights, NY 10598

## Introduction

The need to manage large amounts of data quickly and efficiently is boosting the demand for high-speed data transfer in data centers. The emergence of Generative AI has further fueled the demand for high-speed data transfer such that nearly three-fourths of the data center traffic resides within data centers [1]. The growth in traffic accelerates the need for next-generation networking equipment to support higher port density.  However, traditional copper cables used for data transfer are limited by signal degradation over long distances. This, in turn, drives the requirements for large-scale deployment of high-speed optics to connect the various layers of the networking equipment. Today, data centers rely heavily on optics, but not for the short to mid-length (< 2m) interconnections. The conventional pluggable optics bandwidth increases at a much slower rate than that of data center traffic and the gap between application requirements and the capability of conventional pluggable optics keeps increasing, a trend that is unsustainable.  Co-packaged optics (CPO) is a disruptive approach to increasing the interconnection bandwidth density and energy efficiency by dramatically shortening the electrical link length through advanced packaging and co-optimization of electronics and photonics.

Advances in compute performance have benefitted from Moore's law scaling and have seen up to 60000x in performance in the past 20 years, as shown in Figure 1. However, the I/O bandwidth has increased by a mere 30x in that same timeframe.  The electrical signal rate increase needs to significantly advance to enable the signals to enter/exit and, in addition, there is the concomitant challenge to further move the electrical signals across a card to the front panel of the router or switch, depending on the application.  To solve this challenge, the industry will come to rely on the integration of silicon and optics by co-packaging optical engines along with the main compute chip on a common substrate. This reduces input/output (I/O) power by limiting electrical signaling to intra-package distances, reduces cost by increasing channel count per optical sub-assembly and by eliminating discrete transceiver packaging as well as high-speed PCB traces, and enables high-density optical faceplate connectors.

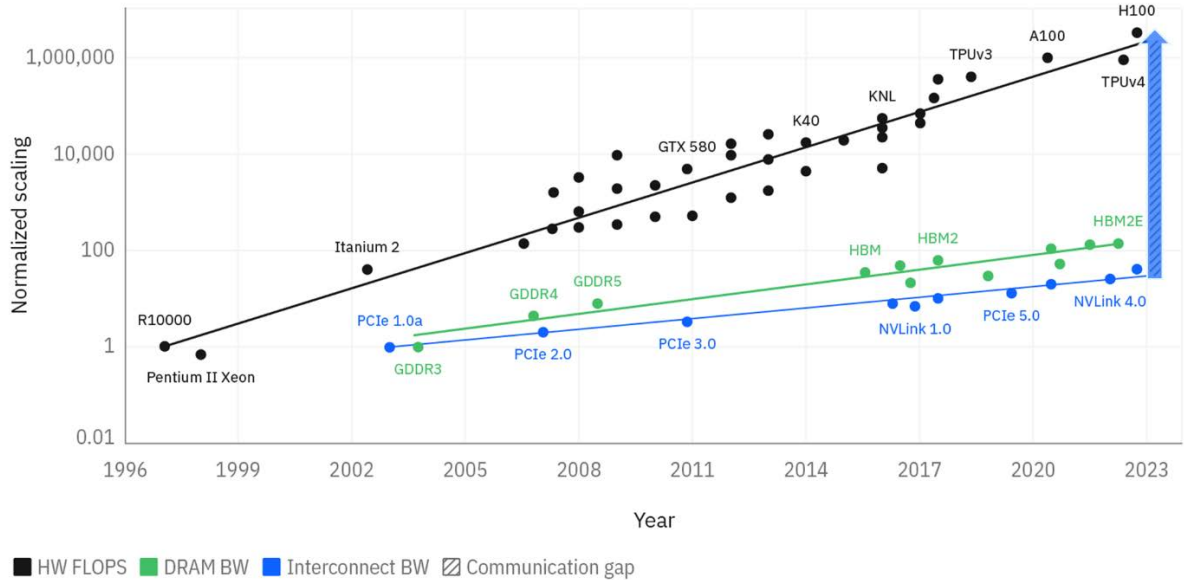

**Figure 1** Scaling of the bandwidth (BW) of different generations of interconnections and memory, as well as the hardware (HW) Peak FLOPS. Note the slow increase in interconnect bandwidth compared to HW FLOPS.

**Training Large Language Models**

Generative AI is transforming society, largely enabled by more efficient hardware, larger datasets, and algorithmic advances, including exponential model size growth that further benefits from more efficient hardware and larger datasets. However, models are scaling rapidly (1000×) and are imposing significant computational demands with increasing model sizes. Developers have thereby relied on distributed training, where multiple accelerators (e.g., GPUs) harness their collective memory capacities and compute capabilities to collaboratively train a large language model (LLM). Distributed training techniques have associated communication between devices to transfer, reduce, aggregate, or exchange information. While communication is necessary, it can limit throughput scaling with increasing device count and cause compute resources to be idle when communication is on the critical path. A recent article [2] showcases that, frequently, networks are the bottleneck when trying to train using GPUs, resulting in one-third of users averaging less than 15% utilization. This causes a significant resource expense, not to mention the significant energy that is spent on training the LLM itself. For models with over a billion parameters, it can take approximately 3 months to train on 8000 H100 GPUs [3]. According to estimates [4], training a GPT-4 model can consume 50 gigawatt-hours of electricity. Moreover, the International Energy Agency calculates that in 2022 data centers consumed 460 terawatt-hours, or almost 2% of global electricity demand, with an expectation that this figure will double by 2026. Further, unlike a system's compute throughput, on which accelerator designers have heavily focused, network bandwidth has not scaled. If compute continues to scale more rapidly, when coupled with increasing communication volume, training future large-scale models on future systems will be inefficient.

To explore the benefits of a CPO solution on model training, we collected data on the training of a llama 3 -70 billion parameter model using a distributed setup with fully sharded data parallel (FSDP) and tensor parallelism (TP). A distributed setup with TP introduces communication and we consider the TP-related communication as the critical path of model execution [5]. For different degrees of TP, we measure the training throughput as shown in Figure 2. In addition, we repeat the measurement for batch sizes of 1 and 2. Our data validates that with increasing TP degree, communication plays an increasingly large role in a distributed training setup and the throughput can be affected as much as 5X. Since batch sizes are typically constrained in training time for improved model quality, TP is usually needed. Increasing the batch size improves the training throughput, however, with increasing degree of TP, the same trends are observed, and throughput is reduced.

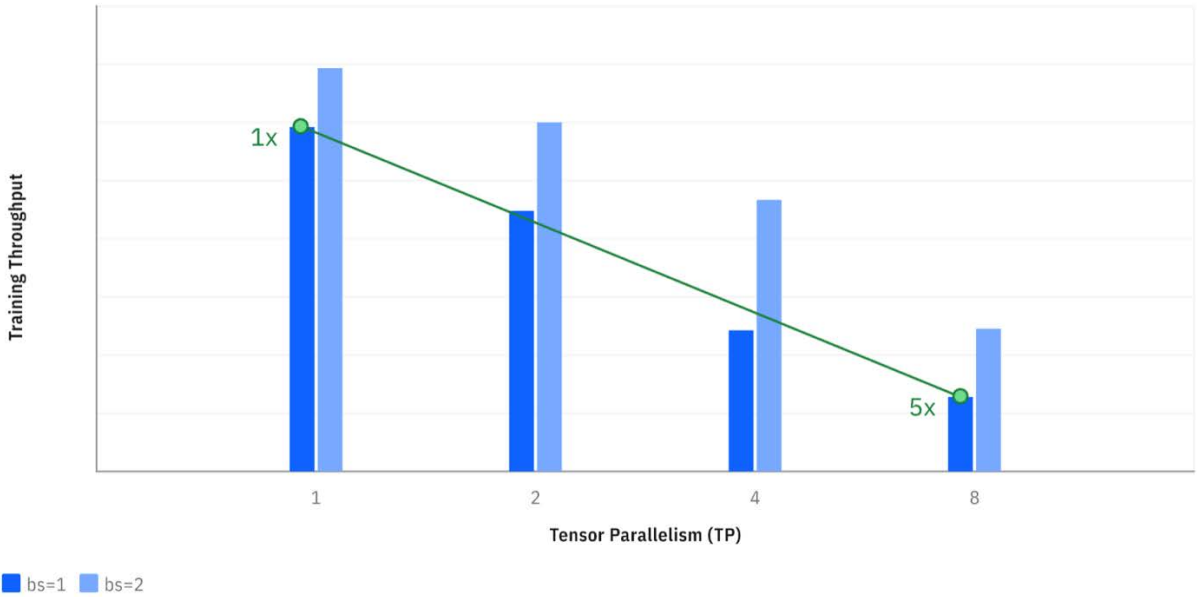

**Figure 2** Training throughput versus degree of Tensor Parallelism (TP) for batch sizes of 1 (bs=1) and 2 (bs=2)

This data showcases that if the communication bottleneck is addressed with CPO, for example, faster AI model training can be achieved. This enables developers to train an LLM up to five times faster with CPO than with conventional electrical wiring. In addition, CPO technology could reduce the time it takes to train a large trillion parameter model from three months [3] to three weeks, with performance gains increasing as the tensor parallel degree and the number of GPUs scale up. The energy savings associated with this gain in training time is significant. Based on the estimates of the power consumption to train GPT-4 [4], the energy savings associated with reducing the communication bottleneck and thereby reducing the training time would be equivalent to powering ~5000 US homes for a year [6].

While Si photonics is not a new concept, advanced fabrication processes and device structures need to be developed to meet the requirements of CPO. CPO solutions have been emerging in recent years. There have been many challenges to the broad adoption of CPO, and increasing the optical fiber integration density on the package is a possible step towards market adoption.

**Co-packaged optics technology**

Prior optical link technologies at IBM have demonstrated use of single-mode fiber (SMF) arrays attach to chip / module with "V" groove attach with 250 $\mu$m pitch and about 127 $\mu$m pitch fiber connection bandwidth [7]. Prior IBM research demonstrations on optical waveguide link samples explored the use of adiabatic coupling for a polymer optical waveguide (PWG) attachment to a photonics integrated circuit chip (PIC) [8, 9] for increased optical link density to a PIC chip and modeling to hardware correlations for optical link budget. Further evaluations and technical publication of polymer optical waveguide to PIC chip optical link only characterization for time zero and post reliability testing to Telcordia – GR-468 testing (A) -4 0C to 85 C (500 cycles), (B) damp heat at 85 C for 85 C / 85 %  relative humidity to 1000 hours and (C) 85 C storage up to 3000 hrs, showed small or no insertion loss changes prior to and post stress test evaluations for most samples under test conditions A and C, whereas most samples under damp heat (B) showed an increase from 0 to 0.75 dB [9]. For PIC chip to polymer optical waveguide to ferrule Telcordia stress testing with conditions (A) -40 C to 85 C (500 cycles), data results for most samples showed an increase for insertion loss from zero to + 0.5 dB [10].

This first-of-a-kind, full-build CPO test vehicle module described herein includes both optical and electrical design elements and a summary of key highlights. Design, modeling, and simulation analysis with correlations to test vehicle hardware for optical link budget are also discussed. This publication will highlight the new co-packaged

optics innovations with combined on module electronics and optics compatibility advancements for:
- (1) module co-design and modeling, photonic, electrical, and integrated CPO module and measurement data
- (2) component design, fabrication and characterization
- (3) Optical test vehicle integration including electronics and optics compatibility for flip chip -lead-free assembly and micro-BGA card attach
- (4) Lid attach and plugable connector to SMF array compatibility, and
- (5) IBM CPO with PWG at 50 um pitch in full module build that passes JEDEC stress testing for manufacturing.

These innovations and/or advanced packaging enhancements support data center bandwidth enhancements and energy reductions for memory to compute and targeted compute to compute chip "all to all" communications bandwidth bottleneck within targeted data centers mid-range to longer range of about 0.5 meters to 10's or 100's of meters.

The key innovations for CPO technology show full module build and characterization with 50 μm pitch optical waveguides to PIC chips with fan-out to ferrules for SMF array attach at 250 μm pitch. The hardware build and characterization data indicate optical waveguides at 18.4 μm to 25 μm pitch can also support low cross talk. Module designs and process flow demonstrations captured optics first and optics last assembly processes. The full module build hardware with lead-free reflow compatibility to support flip chip area array lead-free solder interconnection from chip to substrate, package to board micro- BGA solder attach have passed all JEDEC stress testing results for -40 C to 125 C cycling, damp heat 85C and 85% relative humidity, high-temperature storage at 110 C and low-temperature storage. New materials and structures were required in the project to support each of the test vehicle structures compatibility with optics first assembly process and optics last assembly process, reflow compatibility for PIC to PWG, PWG to the ferrule, module integration, and for JEDEC full module compatibility.

**Module Co-Design & Modeling**

A co-packaged optic module design was developed to support electronic and optics compatibility, industry standards where applicable and scaling for design, process, assembly, test, pluggable connectors, performance / high bandwidth density, lower cost integration and JEDEC stress testing compatibility.

Photonic integrated circuits (PIC) chips were designed and modeled for compatibility with optical waveguides and ferrules as well as base substrates, lids, and cards. PIC,

PWGs, ferrules, and SMFs began with product design kits, where available, to design the optical links for component fabrication and assembly compatibility.   Optical link contributions and projected distributions to insertion loss were modeled and updated with short loop and full build modules to provide test vehicle hardware to optical model updates to improve models and optimize design of experiments (DOE) for efficient learning and hardware to model data and distribution improvements.  Figure 3 shows the preliminary schematic drawings for the optical test vehicle showing a top-down view and side-view representation of the test vehicle components with assembly integration. Here the PIC chip was 8 x 10 mm$^2$, the substrate 17 x 17 mm$^2$ and the optical waveguide was less than 12 mm in length.

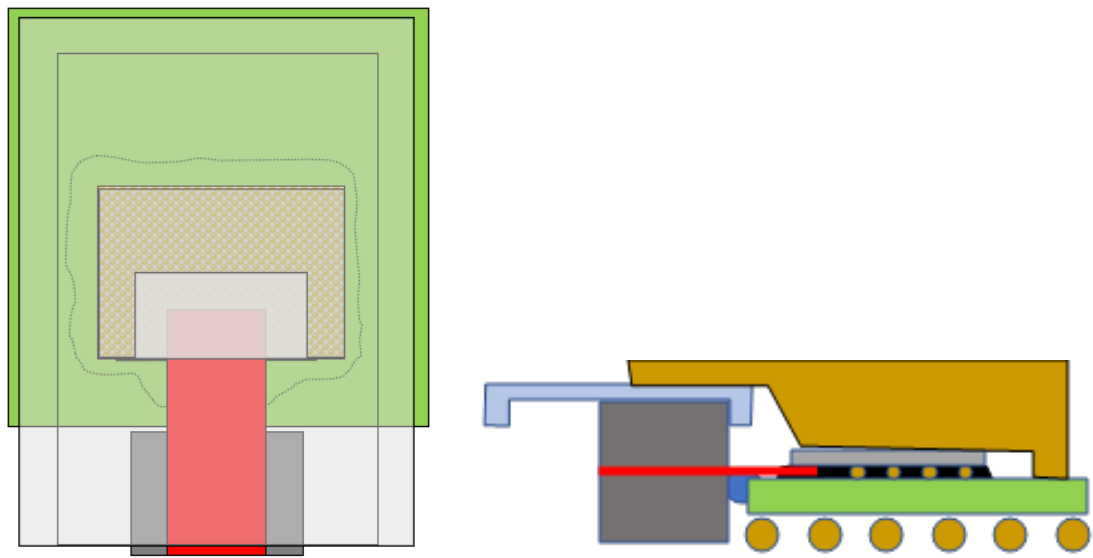

**Figure 3** shows a top down stack view (left) and cross section view (right) of the substrate (green), PIC (grey), optical waveguide (red), ferrule (dark grey), adiabatic coupling area for PIC to PWG (non-bump area), flip chip bump array (small copper colored spheres), micro-BGA array large copper color spheres) and lid (transparent shape top view and copper block shape in cross section, respectively).

The IBM team enhanced optical models and performed simulations through iterations of optical link modeling and correlation to hardware data.  Similarly, use of component and assembly data were used to better understand specification distributions, drive design of experiments (DOE) and improve model and hardware correlations, specifications and validations of model /simulations for OTV-1.  For example, comparison of full optical link insertion loss (IL) targets for each element of the optical link and comparing to component measurements, assembled hardware characterization, improved understanding of the IL losses and reduction in IL through modeling and hardware DOE were made.  In the case of high-density PWG at 50 μm pitch and beginning with PIC chip waveguides which include loop back PIC optical waveguide channels that were adiabatically coupled with PWG to ferrules with pluggable SMFs or arrays, using standard electronic assembly packaging processes. The PIC to SMF optical link models and simulations showed insertion loss could be achieved from under 1.2 to over 2 dB per channel for full link budget pending the parametric and specifications for the assembled link in modules.  Measurements on physical hardware showed that less than 1.2 dB was achieved for some hardware, while a range of 1.5 to 2.0 dB was typical for the channels in assembled OTV-1 hardware with the controlled sample builds within specification components and in specification assembly processes. Some samples had a higher time zero insertion loss, and results showed IP of 1.9 dB to 3 dB loss for post-reflow and JEDEC stress testing.  Opportunities for continuing improvements in IL and quality of samples are expected with time both at time zero and post reflow and post JEDEC stress testing based on root cause analysis and ecosystem collaboration enhancement with suppliers.  In addition, PWG models and simulations were made and compared with build and measurement demonstrations with less than 25 μm pitch.  For example, models / simulations for 18.4 μm PWG pitch have shown less than 30 dB crosstalk in simulations.  Physical measurements for 18.4 μm PWG hardware have verified less than 30 dB cross talk indicating this CPO technology is scalable from 50 μm pitch to <20 μm pitch which can provide very high bandwidth density for chip interconnection at the chip edge.

**Component design, fabrication and characterization**

The IBM photonics team has designed each of the components used in the OTV-1a (optical link first[f] assembly process) and the OTV-1b (optical link last assembly process) and worked with collaborators and either internal and/or external suppliers to expand the ecosystem for prototyping and planning for future production. PIC chiplets, PWGs, ferrules for pluggable fiber connector, industry-available adhesives and custom adhesive properties, substrates, test boards, and lids have been designed, built, and characterized. Vendor data and IBM data characterization comparisons with iterations of build of lots and experimental assessments with characterization have led to improvements in component quality and compatibility for module integration and JEDEC stress testing data/results used to drive necessary component designs, fixturing, specifications, materials, structures, and process improvements.  Next iterations of component enhancements for module assembly compatibility and JEDEC stress test/characterization enhancement results are planned for technical publications in first half 202). Figure 4 shows an example of the PWG for OTV-1 design with 50 μm pitch PIC to PWG adiabatic coupling and 250 μm pitch for PWG to SMF pluggable ferrule attach.

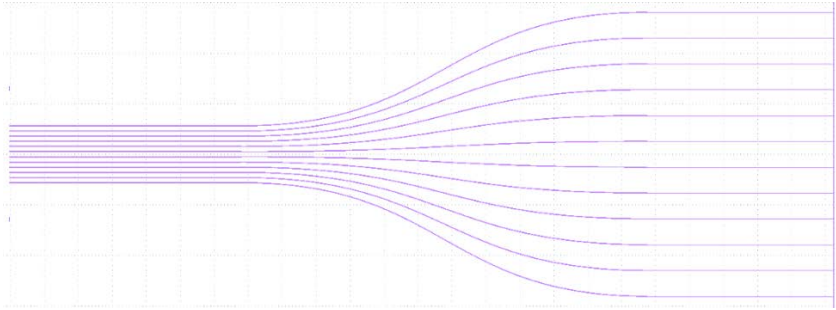

**Figure 4** shows an example of rectangular version of PWG with 50 micrometer pitch waveguide channels fan out to 250 μm pitch waveguide channels.

## Optical Test vehicle integration including electronics & optics compatibility for flip chip -lead free assembly and micro-BGA card attach

Optical test vehicle integration for electronics and optical component assembly compatibility required significant innovations and improvements in component designs, fixturing, specifications, materials and structures combined with process flow and assembly enhancements (updated specifications with controls) to demonstrate flip chip lead-free solder attach and micro-BGA assemblies with minimal or no IL change for assembled modules. Examples of enhancements compared to prior publication included new PIC designs, materials, structures, and specifications, new PWG designs, materials, structures and specifications, new ferrule materials, structures and specifications, new adhesive materials, structures, processes, properties and assembly specifications for PIC attach to PWG and PWG attach to ferrule, new substrates, cards and lid materials, structures, processes and specifications.  Modified processes and control specifications by step were required to demonstrate lead free chip assemblies, micro-BGA attach and PWG assemblies.  Figure 5 shows the schematic drawing of an example PIC, PWG, substrate and ferrule for OTV-1b design, and components and assemblies with PIC to PWG last process.  Figure 3 shows a top view picture of optical test vehicle 1b (OTV-1b) assembled PIC on a cut out substrate and a bottom picture view of PWG with attached ferrule assembled as the last step to PIC to PWG and lid to ferrule.

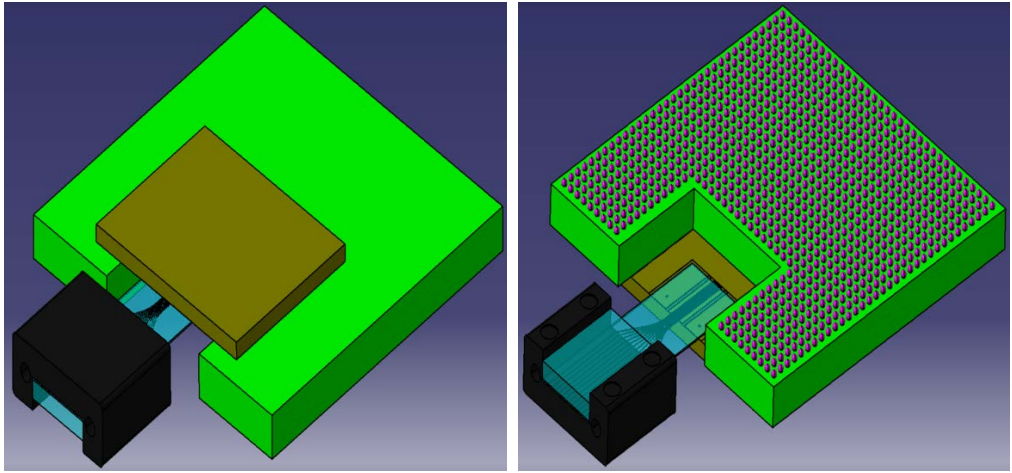

**Figure 5** CAD images of the top view of the PIC, PWG, ferrule and substrate (left) and bottom view of PIC, PWG, ferrule and substrate with micro-BGA and the cut out in the substrate for optics last assembly (right).

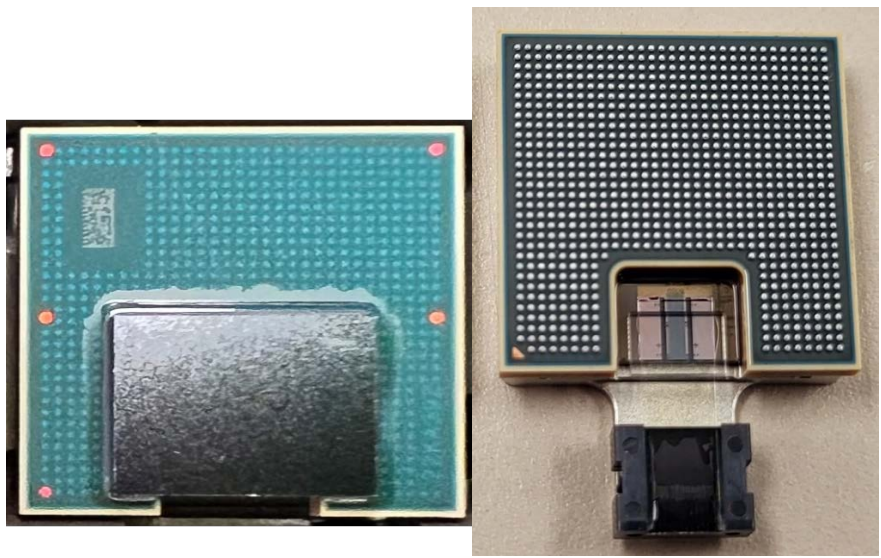

**Figure 6** OTV-1b photos showing top view of PIC to substrate assembly (left) and bottom view picture showing substrate with micro-BGA with PIC and lid attached to top surface and PWG attached to the PIC and LID attached to the ferrule (right) as the last process step in module assembly.

**Lid attach and pluggable connector to SMF array compatibility**

Full module assembly demonstrations were made with OTV-1a (optics to PIC assembly first process flow) and OTV-1b with the optics to PIC assembly last process flow. Each process flow was developed and demonstrated for reflow compatibility for chip attach and micro-BGA assembly, which required components to be compatible with one or more lead-free solder attach processes with no or minimal specification insertion loss change.  In addition to minimal insertion loss change for PIC to PWG in each case the ferrule and SWG to ferrule assembly process and structure needed to be improved for assembly compatibility.  Insertion loss measurements for optical link measurements at no reflow and with 1 to 3 reflow cycles showed samples with 0.0 to + 0.25 dB insertion loss change relative to no reflow cycles for PIC to PWG and PWG to Ferrule assemblies.  Figure 6 shows an example of the assembled module that were characterized for reflow compatibility.

**JEDEC stress testing compatibility**

Full OTV-1A and OTV-1B module builds were made and subjected to multiple reflow / pre-bake preconditioning prior to JEDEC stress testing consisitng of (1) deep thermal cycling of -40 to + 125C  to 1000 cycles, (2) damp heat 85C with 85% relative humidity for 1000 hours, (3) low temperature storage of -40C for 1000 hours and (4) high temperature storage of 110C, 125C or 150C for 1000 hours.   Results showed early samples did not pass these JEDEC stress parametric evaluation from pre-stress test to post stress testing.  Once modeling and associated materials and process enhancements were made to OTV-1A and OTV-1B samples, then the JEDEC stress tests were completed successfully.  Enhancements to processes, adhesives, materials and structures were necessary to improve the results and achieve consistency of samples through /JEDEC testing and characterization.  As part of the continuous improvement enhancements, composition changes have been made and next iterations of evaluation have begun with data and results expected in early 2025.  Pre-testing of samples and characterization indicate the new lots of samples should have improved robustness for JEDEC stress testing with reduced or no change to insertion loss specification targeted at less than 0.0 to + 0.25 dB for pre- versus post JEDEC stress testing.  In addition, current test vehicle hardware demonstrations show greater than 10X improvement in PIC to optical waveguide bandwidth compared to current state-of-the-art (SOA) connectors (See Table 5.1 below).  Note some optical waveguide samples were reduced in width on the PIC side connection to support a higher effective bandwidth density on the PIC to PWG attach and stackable PWG based on mechanical and/or laser micro-cutting.

**Table 1** shows a comparison of current State of the Art (SoA) electrical bandwidth density per second per mm compared to current research demonstrators with benefits from CPO: PIC chip to optical waveguide at 50 μm pitch with one to four wavelength (lambda) compatible links per waveguide.  Future CPO demonstrations with ≤20 μm pitch exhibit low cross talk and future estimates highlight further benefits from new materials, new structures and potential 8 to 16 lambda along with miniaturization of photonics and electronics chips / circuits.

- SOA→ 0.15 to 0.25 Tbps/mm
- CPO→ Hardware demonstrations @ 50 um pitch show 2 to 10 Tbps/ mm
- CPO→ Hardware early measurement & future 4 to 16 lambda opportunity > 20 to 80 Tbps/mm.

**Summary**

The optical test vehicles, OTV-1a and OTV-1b were designed, modeled, fabricated and characterized through lead-free flip chip and BGA assembly processes and through electronic package JEDEC level stress testing.  Results showed that with materials, structures, process, and eco-system collaborations, samples at 50 μm pitch can achieve low insertion loss, passive assemblies with the reflow and JEDEC stress testing evaluations with low or no insertion loss changes.   Modeling and measured data from <18 - 20 μm pitch also indicate that low cross talk can also be achieved indicating that high bandwidth density can be achieved with this multi-channel optical waveguide technology which is a 6X improvement for chip beachfront density compared to IBM's SoA SMF CPO compatible technology.

**Future Work**

The next generation test vehicle with sub 20μm pitch waveguides, increased waveguide channels, increased multiple wavelength (lambda) compatible hardware demonstrations, with options for multiple layer ferrule/connector assembly for the CPO module are currently in development. Modeling and simulation for this future energy efficient hardware demonstration indicate the technology can support increased bandwidth density.  The opportunity to drive enhanced performance with improved energy efficiency for future generative AI  applications and other computing applications is upon us and exciting to drive forward.  Additional detailed CPO technical publications are planned in early 2025.   We look forward to additional performance enhancements, future energy efficiency improvements and an expanding adoption of these technologies for future products.  Compared to current products, these prototypes can support 6X improvement (51 fibers/mm) in PIC chiplet to optical waveguide beachfront density for these test vehicles compared to current SoA fiber connection methods.

**Acknowledgements:**

The IBM CPO team acknowledges the support of ecosystem component suppliers and their respective team members. We also recognize the collaboration across multiple IBM Research, Development and Manufacturing sites, facilities and technical support of the teams who contributed to create these IBM hardware designs, demonstration hardware characterization, assembly / build, test, stress test and diagnostic with characterization.

In addition, the IBM CPO team wishes to recognize the following former IBM contributors to their support during the program: Nicolas Boyer, Qianwen Chen, Nicolas Dupuis, Paul Fortier, Daniel Kuchta and Marie Claude Paquet.